\documentclass{ws-mpla}

\begin{document}

\markboth{H.~Iida}
{PROPERTIES OF SCALAR-QUARK SYSTEMS IN SU(3)$_c$ LATTICE QCD}

\catchline{}{}{}{}{}

\title{
PROPERTIES OF SCALAR-QUARK SYSTEMS \\IN SU(3)$_c$ LATTICE QCD}

\author{\footnotesize HIDEAKI IIDA and TORU T. TAKAHASHI}

\address{Yukawa Institute for Theoretical Physics (YITP), Kyoto University\\
Kitashirakawaoiwake, Sakyo, Kyoto 606-8502, 
Japan\\
iida@yukawa.kyoto-u.ac.jp}

\author{HIDEO SUGANUMA}

\address{Department of Physics, Kyoto University\\
Kitashirakawaoiwake, Sakyo, Kyoto 606-8502, Japan
}

\maketitle

\pub{Received 17 February 2008}{Revised (Day Month Year)}

\begin{abstract}
We perform the first study for the bound states of colored scalar particles $\phi$ (``scalar quarks'') 
in terms of mass generation 
with quenched SU(3)$_c$ lattice QCD. 
We investigate 
the bound states of $\phi$, $\phi^\dagger\phi$ and $\phi\phi\phi$ 
(``scalar-quark hadrons"),
as well as the bound states
 of $\phi$ and quarks $\psi$,
i.e., $\phi^\dagger\psi$, $\psi\psi\phi$ and $\phi\phi\psi$ (``chimera hadrons"). 
All these new-type hadrons including $\phi$ have a large mass of several GeV
due to large quantum corrections by gluons, 
even for zero bare scalar-quark mass $m_\phi=0$ at $a^{-1}\sim 1{\rm GeV}$. We find 
a similar $m_\psi$-dependence between $\phi^\dagger\psi$ and $\phi\phi\psi$, which 
indicates their similar structure due to the large mass of $\phi$. 
From this study, we conjecture that all colored particles generally acquire a large 
effective mass due to dressed gluons. 

\keywords{Dynamical mass generation; Lattice QCD; Scalar-quarks; Diquarks.}
\end{abstract}

\ccode{PACS Nos.: 12.38.Gc, 12.38.Mh, 14.40.Gx, 25.75.Nq}

\section{Introduction}	

The origin of mass is one of the fundamental and fascinating subjects in physics. 
About 99\% of mass of matter in the world originates from 
the strong interaction, which 
provides the large constituent quark mass 
$M_\psi=(300-400){\rm MeV}$.
Such a dynamical fermion-mass generation in the strong interaction can be 
interpreted as spontaneous chiral-symmetry breaking ($\chi$SB).\cite{NJ61,IOS05}

In the strong interaction, however, there is 
other type of dynamical mass generation than $\chi$SB. 
For instance, gluons, which are massless in perturbation QCD, 
seem to have a large effective mass as $(0.5-1.0){\rm GeV}$,\cite{MO87,AS99} 
due to non-perturbative effects.
Actually, glueballs, which are ideally composed only by gluons, 
have a large mass, e.g., about $1.5{\rm GeV}$.\cite{MP99,ISM02} 
The same holds for charm quarks. 
Whereas the current mass of charm quarks is about 1.2GeV
at the renormalization
 point $\mu$ = 1GeV,
the constituent charm-quark mass 
in the quark model is set to be about 1.6GeV. 
The about 400MeV difference between the current and the constituent charm-quark masses 
could be explained by dynamical mass generation without $\chi$SB, 
since there is no chiral symmetry for such heavy quarks. 
These examples imply mass generation without $\chi$SB 
in the strong interaction.
We therefore conjecture that large dynamical mass generation 
generally occurs even without $\chi$SB in the strong-interaction world, i.e., 
{\it all colored particles 
have a large effective mass generated by dressed gluon effects.}\cite{IST07,IST07p}
In this study, we investigate the system of colored scalar particles, 
which do not have chiral symmetry. 

\section{Scalar-quark Hadrons and Chimera Hadrons in Lattice QCD}
We consider light 3c-colored ``scalar-quarks" $\phi$. 
The light scalar-quarks can be also regarded as idealized point-like ``diquarks'' 
at the scale of $a^{-1}\sim 1$GeV. 
We investigate ``scalar-quark mesons'' $\phi^\dagger \phi$ 
and ``scalar-quark baryons'' $\phi\phi\phi$
as the bound states of scalar quarks $\phi$. 
We also investigate the bound states of scalar-quarks $\phi$ and quarks $\psi$, i.e., 
$\phi^\dagger \psi$, $\psi\psi\phi$ and $\phi\phi\psi$, which we name ``chimera hadrons''.\cite{IST07,IST07p}

To include scalar-quarks $\phi$ together with quarks $\psi$ and gluons in QCD,  
we adopt the generalized QCD Lagrangian density,\cite{IST07,IST07p}
\begin{eqnarray}
{\cal L}= -\frac{1}{4}G^a_{\mu\nu}G^{a\mu\nu} + {\cal L}_{\rm F} + {\cal L}_{\rm SQ}, 
\ \ \ {\cal L}_{\rm SQ}={\rm tr} \ (D_\mu \phi)^\dagger(D^\mu\phi)-m_{\phi}^2 \ {\rm tr} 
\ \phi^\dagger\phi,
\end{eqnarray}
where ${\cal L}_{\rm F}$ denotes the quark part and 
$m_\phi$ the bare mass of scalar-quarks $\phi$. 
In the actual calculation, 
we use a discretized Euclidean action on the $16^3 \times 32$ lattice at $\beta$=5.7,\cite{IST07} 
i.e., lattice spacing $a^{-1}\simeq $1.1GeV.\cite{TS0102}
The parameters employed in the analysis are summarized in Table 1. 

\begin{table}[t]
\tbl{The lattice QCD setup for scalar-quark hadrons and chimera hadrons. 
}{
\begin{tabular}{ccccc}
\toprule
$\beta$ &lattice size& lattice spacing $a$&
bare scalar-quark mass $m_{\phi}$ & bare quark mass $m_{\psi}$\\ 
\hline
$5.7$ &$16^3\times 32$&  (1.1GeV)$^{-1} \simeq$ 0.2fm
& 0.00, 0.11, 0.22, 0.33GeV
&0.09, 0.14, 0.19GeV\\
\botrule
\end{tabular}}
\end{table}

\begin{table*}[b]
\tbl{Summary table of new-type hadrons. 
The indices $i,j$ and $k$ denote the scalar-quark 
flavor degrees of freedom. $\Gamma^{ij}_M$, $\Gamma^{ijk}_B$ and 
$\Gamma^{ij}_B$ are some tensors on the scalar-quark flavor.}
{
\begin{tabular}{cccl}\toprule
Names & \ & Lorentz properties &  \ \ \ \ \ \ \ \ Local operators \\
\hline
Scalar-quark meson & ($\phi^\dagger\phi$) & Scalar 
& $M_s(x)= \Gamma^{ij}_M \phi^{\dagger i}_a (x)\phi_a^j(x)$\\
Scalar-quark baryon & ($\phi\phi\phi$)  & Scalar 
& $B_s(x)= \Gamma^{ijk}_B \epsilon_{abc}\phi_a^i (x)\phi_b^j (x)\phi_c^k (x)$\\
Chimera meson & ($\phi^\dagger\psi$) & Spinor 
& $C_M^\alpha(x)= \phi^\dagger_a (x) \psi_a^\alpha (x)$\\
Chimera baryon & ($\psi\psi\phi$) & Scalar 
& $C_{B}(x)= \epsilon_{abc} ({\psi_a}^T(x)C\gamma_5\psi_b(x)) \phi_c (x)$\\
Chimera baryon & ($\phi\phi\psi$) & Spinor & $C_{B}^{\alpha}(x)= 
\Gamma^{ij}_B\epsilon_{abc}\phi_a ^i(x)\phi_b^j (x)\psi_c^\alpha (x)$\\
\botrule
\end{tabular}}
\end{table*}

The gauge-invariant local operators $O(\vec x,t)$ 
of scalar-quark hadrons and chimera hadrons 
are summarized in Table 2. 
We introduce ``scalar-quark flavor" denoted by $i,j,k$ and investigate
the scalar-quark flavor non-singlet mesons, which do not have disconnected
diagrams in their correlators. 
Note that, without the ``scalar-quark flavor'' degrees of freedom, 
the baryonic local operators of $\phi\phi\phi$ and $\phi\phi\psi$ inevitably vanish 
due to the anti-symmetric tensor $\epsilon_{abc}$. 
We calculate the temporal correlator 
$G(t)\equiv \frac{1}{V}\sum_{\vec x} \langle O(\vec x,t) O^{\dagger}(\vec 0,0)\rangle$, 
where the total momentum is projected to be zero.
The mass $M$ of these hadrons are obtained as 
$M \simeq -\frac{1}{T}{\rm ln} G(T)$ for large $T$. 

Here, we show the lattice results for the masses of new-type hadrons. 
Figure 1 shows the squared scalar-quark-meson mass $M_{\phi^\dagger \phi}^2$ 
and the squared scalar-quark-baryon mass squared $M_{\phi\phi\phi}^2$, 
plotted against the bare scalar-quark mass squared $m_{\phi}^2$ at $a^{-1} \simeq 1.1{\rm GeV}$. 
Even for zero bare scalar-quark mass, 
scalar-quark hadrons have a large mass as 
$M_{\phi^\dagger\phi} \simeq 3{\rm GeV}$ and $M_{\phi\phi\phi} \simeq 4.7{\rm GeV}$. 
We find the ``constituent scalar-quark picture'', 
i.e., $M_{\phi^\dagger\phi}\simeq 2M_\phi$ 
and $M_{\phi\phi\phi}\simeq 3M_\phi$, where $M_\phi\simeq (1.5-1.6)$GeV is the
 constituent scalar-quark mass. 
The calculation can be performed 
even in the region $m_\phi^2<0$ due to large quantum corrections on $\phi$.
We also find the relations, 
$M_{\phi^\dagger\phi}^2\simeq 4m_\phi^2+{\rm const.}$ and $M_{\phi\phi\phi}^2\simeq 9m_\phi^2+{\rm const.}$ from the figure. 
Together with the ``constituent scalar-quark picture'', we reach the relation 
$M_\phi^2\simeq m_\phi^2+\Sigma_\phi$, where $\Sigma_\phi$ is the self-energy of $\phi$ and 
is expected to be 
insensitive to $m_\phi$. 
This is a natural relation between the renormalized mass and the bare 
mass for scalar particles.

\begin{figure}[t]
\begin{center}
\includegraphics[width=10cm]{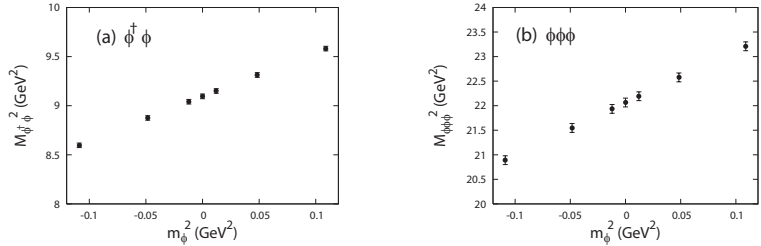}
\caption{
(a) ``Scalar-quark meson'' mass squared $M_{\phi^\dagger\phi}^2$ and 
(b) ``scalar-quark baryon'' mass squared $M_{\phi\phi\phi}^2$ 
plotted against the bare scalar-quark mass squared $m_{\phi}^2$ 
at $a^{-1} \simeq 1.1{\rm GeV}$.}
\label{fig2}
\end{center}
\end{figure}

\begin{figure}[b]
\begin{center}
\includegraphics[width=6.1cm]{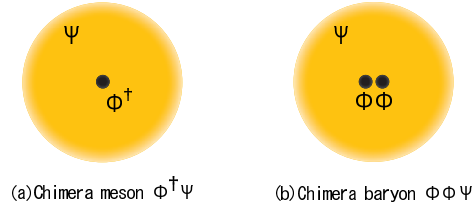}
\caption{
The conjectured wave-function of the quark $\psi$ 
around the heavy scalar-quarks in the chimera hadrons. 
Similar to chimera mesons, 
the wave-function of $\psi$ in chimera baryons distributes 
around the two $\phi$'s localized near the center of mass.
}
\label{fig2}
\end{center}
\end{figure}

Chimera hadrons also have a large mass even at $m_\phi=m_\psi=0$, i.e., 
$M_{\phi^\dagger\psi}\simeq 1.9$GeV for chimera mesons $\phi^\dagger\psi$, and 
$M_{\psi\psi\phi}\simeq 2.2$GeV, $M_{\phi\phi\psi}\simeq 3.6$GeV for 
chimera baryons ($\psi\psi\phi, \phi\phi\psi$). 
We find a ``constituent scalar-quark/quark picture'', i.e., 
an approximate relation as $M_{m \phi +n \psi} \simeq m M_{\phi} + n M_{\psi}$  
with the constituent quark mass $M_{\psi}\simeq 0.4{\rm GeV}$, and the
 large constituent scalar-quark mass $M_{\phi}\simeq (1.5-1.6){\rm GeV}$.\cite{IST07} 

From the $m_\psi$-dependence of chimera hadron masses, 
we conjecture a similar structure between chimera mesons $\phi^\dagger\psi$ and 
 chimera baryons $\phi\phi\psi$.\cite{IST07} 
The wave-function of $\psi$ in a chimera meson $\phi^\dagger\psi$ is distributed 
around the heavy scalar-quark $\phi^\dagger$ due to the large mass of $\phi$, 
and, similarly, the wave-function of $\psi$ in a chimera baryon $\phi\phi\psi$ is 
distributed around the point-like ``di-scalar-quark" $\phi\phi$. (See Fig.~\ref{fig2}.)

\section{Summary and Conclusion}
We have performed the first study of light ``scalar-quarks" $\phi$ 
(colored scalar particles or idealized diquarks) 
and their color-singlet hadronic states in quenched SU(3)$_c$ lattice QCD 
in terms of dynamical mass generation. 
We have investigated the mass of ``scalar-quark mesons'' $\phi^\dagger \phi$, 
``scalar-quark baryons'' $\phi\phi\phi$ and ``chimera hadrons''  
($\phi^\dagger \psi$, $\psi\psi\phi$, $\phi\phi\psi$), 
which are composed of quarks $\psi$ and scalar-quarks $\phi$. 
We have observed the large dynamical mass generation of scalar-quarks $\phi$ about 1.5GeV 
at $a^{-1}\simeq 1.1{\rm GeV}$ due to large quantum corrections by gluons, 
even at the zero bare scalar-quark mass $m_\phi=0$.
This lattice result also indicates that plausible diquarks used in effective hadron models 
cannot be described as the point-like particles and 
should have a much larger size than 
$a\simeq 0.2$fm.\cite{IST07,IST07p}

This study indicates that, even without $\chi$SB, 
large dynamical mass generation in the strong interaction occurs for the scalar-quark systems. 
Together with the large glueball mass and 
the large difference between the current and the constituent charm-quark masses, 
this type of mass generation would generally occur in the strong interaction, 
and therefore we conjecture that all colored particles generally acquire 
a large effective mass due to dressed gluon effects,\cite{IST07,IST07p} as shown in Fig.~3.
\begin{figure}[ht]
\begin{center}
\includegraphics[width=8cm]{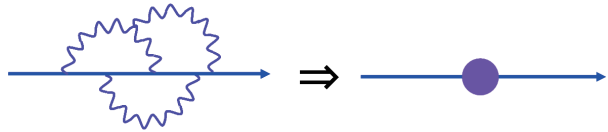}
\caption{Schematic figure for dynamical mass generation of colored particles. 
Even without chiral symmetry breaking, colored particles generally acquire 
a large effective mass due to dressed gluons.
}
\label{fig3}
\end{center}
\vspace{-0.65cm}
\end{figure}

\end{document}